\documentclass[aps, prd, twocolumn, showpacs, superscriptaddress, groupedaddress]{revtex4}

\usepackage{graphicx}	
\usepackage{amssymb}
\usepackage{multirow}
\usepackage{dcolumn}
\usepackage{color}
\usepackage{amsmath}
\usepackage{comment}
\usepackage{subfigure, rotating, bm, array}
\usepackage[pagebackref=false, colorlinks=true]{hyperref}
\hypersetup{
linkcolor=blue,     
citecolor=blue,     
urlcolor=blue}      
\begin{document}
\title{Observational aspects of a class of Dark matter spacetimes}
\author{Ashok B. Joshi}
\email{gen.rel.joshi@gmail.com}
\affiliation{International Center for Cosmology, Charusat University, Anand, GUJ 388421,  India}
\author{Divya Tahelyani}
\email{tahelyanidivya118@gmail.com}
\affiliation{International Center for Cosmology, Charusat University, Anand, GUJ 388421, India}
\author{Dipanjan Dey}
\email{deydipanjan7@gmail.com}
\affiliation{Department of Mathematics and Statistics, Dalhousie University, Halifax, Nova
Scotia, Canada, B3H 3J5}
\author{Pankaj S. Joshi}
\email{psjcosmos@gmail.com}
\affiliation{Cosmology Centre, Ahmedabad University, Ahmedabad, GUJ 380009, India}
\affiliation{International Center for Cosmology, Charusat University, Anand, GUJ 388421,  India}
\date{\today}
\begin{abstract}
Various astrophysical and cosmological observations today serve as an indirect evidence of the existence of dark matter in the universe. In the present work, we propose a class of spacetimes that show some important characteristics relevant to the spacetime of a dark matter halo.  These spacetimes are static and spherically symmetric solutions of the Einstein field equations. The proposed spacetimes satisfy the flat velocity profile of a galactic object far away from the center, they give good agreement to the astrometric data of the S2 star, and these also cast a central shadow. Using the Penrose diagram, we show that the causal nature of the central singularity here is null. 
This single spacetime model for galactic dark matter may therefore be used to explain some of the properties of galactic dynamics at different length scales.
\bigskip

$\boldsymbol{key words}$ : Dark matter, Naked singularity spacetime, Black hole spacetime.

\end{abstract}
\maketitle
\section{introduction}\label{section1}
One source of indirect evidence for the existence of dark matter comes from observations of the rotation curves of galaxies. The rotation curves of the spiral galaxies demonstrate that the velocity of the stars and gas in the outer regions of these galaxies becomes almost constant~\cite{Persic:1995ru}. Therefore, their rotation curves are flat, in contrast to what is expected from Newtonian dynamics. The most general explanation is the presence of an additional, unseen mass component that causes the observed motion. The galactic rotation curves are generally divided into three parts: a bulge, a disk, and a dark halo. 
In general, the matter distribution in our galaxy is classified into four different parts. The radial distance from $10^{-4}$pc to $2$pc consists of young stars which follow Keplerian law\,($v \propto r^{-1/2}$)~\cite{Krut:2018ygt}. Non-circular motions of young stars have frequently been observed near the galactic center. An intermediate spheroidal bulge structure comprises older stars that follow the exponential spheroidal model and ranges from $2$pc to $10^3$pc. The region $10^3 - 10^4$pc is called an extended flat disk in which dust and gas are present, which is a star-formation zone. The region $10^4 - 10^5$pc is dominated by dark matter called a spherical or outer halo~\cite{Krut:2018ygt}.

The nature of dark matter remains very much unknown as of now. There are many possible candidates for dark matter, such as heavy neutrinos, Weakly Interacting Massive particles (WIMP), micro and primordial black holes, etc~\cite{dmcandidate}, and others. As an alternative to dark matter, several theoretical models, based on a modification of Newton’s law or of general relativity, have been proposed so far to explain the behaviour of the galactic rotation curves~\cite{Milgrom:1983ca, Mannheim:1996rv, Roberts:2002ei, Bekenstein:2004ne, Moffat:1995dq, Brownstein:2005zz, Mak:2004hv, Harko:2005yk, Boehmer:2007xh}. Till now, there has been no complete theory that predicts the direct detection of dark matter.
Our Milky Way galaxy serves as an experimental hub to test general relativity on the largest as well as the smallest scale. Sagittarius A* (Sgr A*) is a massive and compact radio source located at the center of our galaxy. One of the primary concerns about Sgr A* is its true identity. It is still a mystery whether it is a black hole or an alternative, such as a naked singularity, a wormhole, or any other exotic compact object. Many investigations into gravitational collapse have been conducted, and it has been discovered that the final state of collapse is not necessarily a black hole. There is a considerable amount of research showing that null and timelike singularities can be formed during the gravitational collapse of physically plausible matter clouds~\cite{Joshi:1993zg, Joshi:2011zm, Mosani:2020ena, Dafermos:2017dbw, Bhattacharya:2017chr}, which contradicts the Cosmic Censorship Conjecture (CCC) proposed by Rodger Penrose~\cite{RPenrose}. The galactic center with a strong gravity region can be explored by investigating numerous observable factors, such as the shadow of the ultra-compact object, the relativistic orbit of the S2 star, accretion disk features, etc. In 2019, the Event Horizon Telescope (EHT) released the first groundbreaking horizon-scale image for M87~\cite{EventHorizonTelescope:2019dse}, a supermassive compact object present at the center of galaxy Messier 87. This was followed by the recent release of the first image of Sgr A*~\cite{EventHorizonTelescope:2022xnr, EventHorizonTelescope:2022tzy, EventHorizonTelescope:2022ago, EventHorizonTelescope:2022vjs, EventHorizonTelescope:2022wok, EventHorizonTelescope:2022urf, EventHorizonTelescope:2022exc, EventHorizonTelescope:2022gsd, EventHorizonTelescope:2022xqj}. To know the actual theoretical framework of the geometry around the Sgr A* compact object, many groups are working on the shadow and accretion disk properties of the compact object~\cite{Joshi2020,Dey:2020haf,Paul2020,Dey:2020bgo,Shaikh:2018lcc,Tahelyani:2022uxw, Liu:2020vkh, Liu:2021yev, Joshi:2013dva, Bambhaniya:2021ugr, Rahaman:2021kge, Harko:2008vy, Harko:2009xf, Kovacs:2010xm, Harko:2009gc, Guo:2020tgv, Chowdhury:2011aa}. On the other hand, decades of continuous astrometric and spectroscopic observations by GRAVITY, SINFONI, and the UCLA Galactic Center group provide useful information about the central compact object by using the trajectories of `S'-stars. These `S'-stars are located very close to the galactic center and are orbiting around Sgr-A* with a very high velocity ~\cite{datasupplementary, GRAVITY:2018ofz, Hees:2017aal, GRAVITY:2020gka}. In this context, there is a large amount of literature in which the nature of the timelike trajectories and orbital precession of a test particle in various spacetime geometries is extensively studied~\cite{Martinez, Eva,Eva1,Eva2,tsirulev,Joshi:2019rdo,Bambhaniya:2019pbr,Dey:2019fpv,Bam2020,Lin:2021noq,Deng:2020yfm,Deng:2020hxw,Gao:2020wjz,aa4,Glampedakis:2002ya,Fujita:2009bp,Pugliese:2013zma,rana}.

In the present work, we propose a new procedure to model the spacetime of galactic dark matter halo in the framework of general relativity. We construct a class of spacetimes here that can model the flat rotation curves far from the center, and these have a photon sphere near the center. First, we use the well-established observational constancy of the rotation curves far away from the galactic center. Further, we use the conditions for the presence of a photon sphere in spacetime. These properties allow us to obtain the metric components for the spacetime we construct here. The available observational data of the astrometric positions of the S2 star (one of the stars in the `S'-stars' family) can provide useful constraints on the proposed model. We obtain the timelike orbit of the massive particle in the proposed dark matter spacetime and constrain the free parameters of the metric by fitting the theoretical orbit with the observational data of the S2 star orbit. Furthermore, we study the rotation curve, redshift, and shadow properties using the best-fitted parameters. 

The paper is organized as follows: In Section (\ref{section2}), we construct the class of spacetimes using the properties of the flat velocity curve and photon sphere. We also discuss the energy conditions and causal structure of spacetime in that section. Subsequently, in Section~(\ref{section3}), we study the particle trajectory in the proposed dark matter spacetime, and after that, we obtain the best-fitted parameters' values using the orbital motion of the S2 star. Next, in Section~(\ref{section4}) and Section~(\ref{section5}), we study the circular velocity profiles far away from the center, and the shadow properties of the spacetime using the best fitted values of parameters. Finally, in Section~(\ref{section6}), we conclude our results.


\section{Construction of dark matter spacetime}\label{section2}
Observations show that almost $90\%$ of the galaxy should be made up of dark matter. Therefore, at a considerable distance from the center, we can assume that compared to dark matter, baryonic matter does not contribute significantly to the total energy density of the halo and hence to the dynamics of particles in the galaxy. Therefore, after a certain distance, we can assume that the luminous matter acts as a test fluid that travels in the curvature created by the dark matter.

To begin with, we assume that stars behave as test particles that follow the timelike geodesics of a static and spherically symmetric spacetime. In this situation, the most generic form of the line element of spacetime is as follows:
\begin{equation}
    ds^2 = - g_{tt} c^2 dt^2 + g_{rr}dr^2 + r^2(d\theta^2 + \sin^2\theta d\phi^2)\,\,,
    \label{metric}
\end{equation}
where $g_{tt}$,~$g_{rr}$ are the functions of $r$ only, and the azimuthal part of the spacetime exhibits the spherical symmetry. In this section, we intend to derive a spacetime metric that possesses two properties: (i) a flat velocity curve away from the galactic center and (ii) a photon sphere near the galactic center.

The rotation curves of spiral galaxies are one of the most important indirect evidence of the existence of dark matter. In the spiral arms of theses galaxies, neutral hydrogen (HI) clouds host millions of stars. The frequency shifts in these clouds' 21 cm HI emission are used to measure the velocities of the stars which are hosted by the cloud. It is an empirical fact that, far from the galactic center, the celestial bodies follow almost circular orbits and the circular velocity of stars is a frame-dependent quantity. The shift in 21 cm HI emission occurrs due to the Doppler effect caused by local Lorentzian motion of the stars. Therefore, to investigate the possible metric components which admit flat velocity profile far away from the center, one needs to use the expression of the circular velocity of a particle measured by a stationary Lorentzian observer at one location in the orbit. The Lagrangian for the test particle travelling in any spacetime can be written as
\begin{equation}
    2\mathcal{L} = g_{\mu \nu}\frac{dx^{\mu}}{ds}\frac{dx^{\nu}}{ds},  \label{L1}
\end{equation}
where $s$ is the affine parameter along the geodesic. For timelike geodesics, the affine parameter is the same as the proper time $\tau$ of the particle. For the sake of simplicity, we constrain our analysis to the equatorial plane ($\theta=\pi/2$). For the line element described in Eq.~(\ref{metric}), it follows that
\begin{eqnarray}
      2\mathcal{L} &=& - g_{tt} \dot{t}^2 + g_{rr} \dot{r}^2 + r^2 \dot{\theta}^2 + r^2 \sin^2 \theta \dot{\phi}^2\,.\nonumber\\
      &=&\epsilon
\end{eqnarray}
Here, the dot indicates the derivative with respect to the proper time $\tau$. The null and timelike geodesics are characterised by $\epsilon=0$ and $\epsilon=-c^2$, respectively. The conserved quantities, energy~($e$) and angular momentum~($h$) of the particle per unit rest mass can be obtained as follows:
\begin{eqnarray}
    -e & = &\frac{\partial \mathcal{L}}{\partial \dot{t}}=-g_{tt} c^2 \dot{t}\label{e}\,,\\
    h & = &\frac{\partial \mathcal{L}}{\partial \dot{\phi}} = r^2 \dot{\phi} \label{h}\,.
\end{eqnarray}
Since for timelike case $2 \mathcal{L}=-c^2$, eliminating $\dot{t}$ and $\dot{\phi}$ using Eqs.~(\ref{e}) and (\ref{h}), we obtain
\begin{equation}
   E = e^2 c^2 = g_{tt} g_{rr} \dot{r}^2 + V_{eff}\,,\label{E}
\end{equation}
and the effective potential
\begin{equation}
    V_{eff}(r) = g_{tt} \left(c^2 + \frac{h^2}{r^2}\right)\,.\label{vr}
\end{equation}
Here, $E$ is the total energy of the particle. Since we are interested in the stable circular orbit of the test particle, the following conditions must be satisfied:
\begin{eqnarray}
   &\text{(a)}& \dot{r} = 0\,\,\text{(circular orbit)}\,,\\
   &\text{(b)}&  \frac{dV_{eff}(r)}{dr} =0\,\,\text{(extremum motion)},\\
   &\text{(c)}& \frac{d^2 V_{eff}(r)}{dr^2}\bigg|_{extrem} > 0\,\,\text{(stable orbit)}.
\end{eqnarray}
Solving the conditions (a) and (b) for $e$ and $h$, we obtain
\begin{eqnarray}
    e &=& \sqrt{\frac{2 g_{tt}^2}{2 g_{tt}-r g'_{tt}}}\,, \\
    h &=& c\sqrt{\frac{r^3 g'_{tt}}{2 g_{tt}-r g'_{tt}}}\,,
\end{eqnarray}
where $(')$ denotes derivative with respect to the radial coordinate $r$. By substituting the above expressions of $e$ and $h$ in Eqs.~(\ref{e}) and (\ref{h}), the four-velocity of the particle becomes
\begin{equation}
    u^{t} = \dot{t}= \frac{1}{c^2}\sqrt{\frac{2}{2 g_{tt}-r g'_{tt}}},\,\,\,
    u^{\phi} = \dot{\phi} = c\sqrt{\frac{g'_{tt}}{r(2 g_{tt}-r g'_{tt})}}\,.\label{ut}
\end{equation}
While moving along orbit, when the position of the particle coincides with the position of the observer, the velocity of the particle can be measured at that instant of time by using the local Lorentz tetrad frame of the observer. The basis vectors of the orthonormal tetrad frame of the stationary observer can be written as~\cite{hartle}
\begin{eqnarray}
    e^{\alpha}_{\hat{t}} &=& \left\{\frac{1}{\sqrt{c^2 g_{tt}}}, 0, 0,0\right\}\,,\\
    e^{\alpha}_{\hat{r}} &=& \left\{0,\frac{1}{\sqrt{g_{rr}}}, 0,0\right\}\,,\\
    e^{\alpha}_{\hat{\theta}} &=& \left\{0, 0, \frac{1}{r},0\right\}\,,\\
    e^{\alpha}_{\hat{\phi}} &=& \left\{0, 0, 0, \frac{1}{r \sin{\theta}}\right\}\,\,,
\end{eqnarray}
where $(\,\hat{}\,)$ corresponds to the coordinates of the orthonormal tetrad frame. Using the above basis vectors, the $t$ and $\phi$ components of the four velocities in the observer's local frame can be obtained
\begin{eqnarray}
    u^{\hat{t}}=\frac{1}{c^2}\sqrt{\frac{2 g_{tt}}{2 g_{tt}-rg'_{tt}}},\,\,\,
    u^{\hat{\phi}} &=& c\sqrt{\frac{r g'_{tt}}{2 g_{tt}-r g'_{tt}}}\,.
\end{eqnarray}
The tangential velocity~($v$) of the particle in the circular orbit as measured by the Lorentzian observer can be defined as
\begin{equation}
    v = v^{\hat{\phi}}= \frac{u^{\hat{\phi}}}{u^{\hat{t}}}= c \sqrt{\frac{rg_{tt} ^{'}(r)}{2g_{tt} (r)}} \,\,.\label{velocity}
\end{equation}

We know that the observations in the spiral galaxy suggest that very far from the galactic center, $v\approx constant$~\cite{Rubin:1980zd}. Using this condition, the integration of the above equation gives the functional form of the $g_{tt}(r)$, which we denote as $g_{tt_1}(r)$
\begin{equation}
   g_{tt_1}(r) = e^{\left(\frac{2V_{c}^2}{c^2}\right) \ln\left(\frac{r}{R_c}\right)}, \label{gtt1}
\end{equation}
where $V_{c}$ is the constant velocity of the stars far away from the center and $R_c$ is the constant of integration.
The functional form Eq.~(\ref{gtt1}) of $g_{tt}$ is valid only in the region of the flat rotation curve. To find the condition on $g_{tt}$ in the region close to the center, we use the aforementioned second property of the metric: the presence of a photon sphere. We know that the photon sphere corresponds to the unstable circular orbits of the photons, and its presence in spacetime requires the existence of a maximum of the effective potential ($\mathcal{V}_{eff}$) of null geodesics. 
For the spacetime described in Eq.~(\ref{metric}), in the equatorial plane~($\theta=\pi/2$), the null geodesics satisfies the relation\begin{equation}
    \frac{1}{b^2}=\frac{g_{tt} g_{rr}}{l^2} \left(\frac{dr}{d\lambda}\right)^2 + \mathcal{V}_{eff}\,\, ,
\end{equation}
where the effective potential 
\begin{equation} \label{veff}
    \mathcal{V}_{eff} = \frac{g_{tt}}{r^2}\,, 
\end{equation}
and the impact parameter $b = \frac{l}{\varepsilon}$, where $\varepsilon$ and $l$ are conserved energy and conserved angular momentum of the photon respectively. 

At photon sphere radius~($r_{ph}$), one can write $\mathcal{V}_{eff}(r_{ph})=\frac{\varepsilon^2}{l^2},\, \mathcal{V}_{eff}'(r_{ph})=0 \text{, and } \mathcal{V}_{eff}''(r_{ph})<0$. In order to obtain the spacetime metric that possesses a photon sphere, we start by considering an ansatz for the effective potential, which has a maximum as follows:
\begin{equation}\label{veff1}
    \mathcal{V}_{eff}=\frac{1}{r^2}e^{p-(\frac{r_{*}}{r})^n}\,,
\end{equation}
where $p$, and $r_{*}$ are constants. Using the condition $\mathcal{V}_{eff}'(r_{ph})=0$, we can calculate the radius of photon sphere as
\begin{equation}\label{rph}
    r_{ph}= \left(\frac{2}{n}\right)^{-\frac{1}{n}} r_{*}\,.
\end{equation}
Substituting the expression of $r_{*}$ from the above equation in the Eq.~(\ref{veff1}), the form of effective potential becomes
\begin{equation}
    \mathcal{V}_{eff}= \frac{1}{r^2}e^{p-\frac{2}{n}(\frac{r_{ph}}{r})^n}\,.
\end{equation}
 From Eq.~(\ref{veff}), we can write the second form of $g_{tt}(r)$~(i.e., $g_{tt_2}$)
\begin{equation} \label{gtt2}
   g_{tt_2} = e^{p-\frac{2}{n}(\frac{r_{ph}}{r})^n}\,.
\end{equation}
For the above type of form of $g_{tt}$, the maxima and upper-bound of the effective potential coincide which allows the spacetime to cast a shadow \cite{Joshi2020, Dey:2020bgo}. 
The functional form of $g_{tt}$ in Eq.~(\ref{gtt1}) becomes dominant as one moves away from the center, accounting for the particle's constant velocity in the outer parts of the galaxy. On the other hand, Eq.~(\ref{gtt2}) dominates over small distances from the center and creates the photon sphere near the center. Now, Eqs.~(\ref{gtt1}) and (\ref{gtt2}) can be combined to obtain a single expression of $g_{tt}$ that exhibits both properties: a presence of a photon sphere near the center as well as a flat velocity profile farther away from the center.
\begin{equation} \label{gtt3}
g_{tt}(r)= q*e^{\left(p - 2(\frac{c^2-V_c^2}{c^2 n})\left(\frac{r_{ph}}{r}\right)^n\right) + \frac{2V_c^2}{c^2}\ln\left(\frac{r}{R_c}\right)}\,.
\end{equation}
where, $q$ is a constant that can be determined from the boundary conditions. Here, Eqs.~(\ref{gtt1}) and~(\ref{gtt2}) are combined in such a way that the photon sphere still remains at the same radius  $r=r_{ph}$ in the merged expression of $g_{tt}$. This spacetime of the galactic dark matter halo cannot be an asymptotically flat spacetime. It should be matched with any asymptotically flat spacetime at some timelike hypersurface. 

In general relativity, for the smooth matching of two spacetimes at a timelike or spacelike hypersurface, two junction conditions must be fulfilled on the matching hypersurface~\cite{Poisson:2009pwt}. The first condition requires that the induced metric~($h_{ab}$) on both sides of the matching hypersurface must be identical, and the second condition states that the extrinsic curvature~($K_{ab}$) of the internal and external spacetimes at the matching hypersurface should be the same. The extrinsic curvature can be written in terms of the covariant derivative of normal vectors on the hypersurface:
\begin{equation}
     K_{ab} = e^{\alpha}_a e^{\beta}_{b}\nabla_{\alpha}\eta_{\beta}\,,
\end{equation}
where, $e^{\alpha}_a$ is the tangents to the hypersurface and $\eta^{\beta}$ is the normal to that hypersurface.
We suppose that the spacetime geometry outside the halo is described by the Schwarzschild spacetime in the weak field limit, which joins with the internal dark matter geometry at the matching hypersurface $r=R_b$. The line element of exterior Schwarzschild spacetime in the weak field limit can be written as 
\begin{equation}\label{ext_metric}
    ds_{ext}^2 = - c^2 (1 -\alpha) dt^2 + (1 +\alpha) dr^2 + r^2d\Omega^2\,\, ,
\end{equation}
where, $\alpha = \frac{2V_{b}^2}{c^2}$ and $d\Omega^2=d\theta^2+\sin^2\theta d\phi^2$.
Here, $V_b$ is the velocity of the particle, which would be the same as the constant velocity of the stars in the flat velocity profile region. It should be noted that $R_{b}>>r_{ph}$. One can check that the spacetime geometry comprising the internal and external metrics is not smooth at the junction hypersurface $r=R_{b}$. Therefore, there must be a thin shell of matter at the junction. Using the induced metrics matching at the junction, we get $q=c^2$ and $p=-\alpha$. 
By considering the tangential velocity $V_c=V_b$ at $r=R_b$, we can determine the integration constant $R_c$ in Eq.~(\ref{gtt1}). Thus, Eq.~(\ref{gtt1}) becomes
\begin{equation}
   g_{tt_1}(r) = e^{\left(\frac{2V_{b}^2}{c^2}\right) \ln\left(\frac{r}{R_b}\right)}\,. 
\end{equation}
The final temporal component of the internal metric $g_{tt}(r)$ can be written as
\begin{equation}
g_{tt}(r)= c^2 Exp\left[{ -\alpha - \frac{\beta(2-\alpha)}{n} + \alpha\log\left(\frac{r}{R_b}\right)}\right]
\label{gtt} 
\end{equation}
where, $\beta = \left(\frac{r_{ph}}{r}\right)^n$.

We now address the question of how to obtain $g_{rr}(r)$. We proceed by solving Einstein's field equations to determine $g_{rr}$ inside the halo. We consider that the dark matter that comprises the spherical halo is a fluid with the energy density, $c^2\rho(r)$; radial pressure, $P_r(r)$; and tangential pressures, $P_{\theta}(r)$ and $P_{\phi}(r)$. For the matching of the internal metric with the external Schwarzschild metric on the matching hypersurface, radial pressure at the matching boundary of the internal spacetime should be zero. For simplicity, we consider that the radial pressure inside the internal spacetime is throughout zero.
Now, one can solve Einstein's field equations with the condition $T_{1}^{1} = P_{r} = 0$ using previously determined $g_{tt}(r)$~(Eq.~(\ref{gtt})), and deduce the expression for $g_{rr}(r)$ as 
\begin{equation}
g_{rr}(r)= (1 + 2\beta + \alpha (1 - \beta))\,.
\label{grr} 
\end{equation}
By substituting $g_{tt}(r)$ and $g_{rr}(r)$ in Eq.~(\ref{metric}), the line element of the proposed dark matter spacetime becomes
\begin{widetext}
\begin{eqnarray}
ds_{int}^2 = - c^2 Exp\left[{ -\alpha - \frac{(2-\alpha)}{n}\left(\frac{r_{ph}}{r}\right)^n + \alpha\log\left(\frac{r}{R_b}\right)}\right] dt^2 + \left(1 + 2\left(\frac{r_{ph}}{r}\right)^n + \alpha \left(1 - \left(\frac{r_{ph}}{r}\right)^n \right)\right)dr^2 + r^2d\Omega^2\, . \label{dm_metric}
\end{eqnarray}
\end{widetext}
 Note that the Eq.~(\ref{dm_metric}) represents a class of spacetimes that depend on the parameter $n$. One can also select a different ansatz for the effective potential and obtain another class of spacetimes.
\subsection{Causal structure of the proposed Dark matter spacetime and Energy conditions}
From the expressions of the Kretschmann scalar\,($ R_{\alpha\beta\gamma\delta}R^{\alpha\beta\gamma\delta}$) and Ricci scalar, it can be seen that both the scalars diverge at $r=0$ which implies the existence spacetime singularity at the center.

To have an event horizon in a static spacetime, the following condition must be followed at the event horizon radius ($r_e$)
\begin{equation}
    \lim_{r\to r_e} \frac{dt}{dr} = \pm \sqrt{\frac{g_{rr}}{g_{tt}}}= \pm \infty. \label{eq11}
\end{equation}
One can check that in the proposed dark matter spacetime (Eq.~\ref{dm_metric}), the above condition is satisfied only for $r_{e}=0$. This implies that the spacetime does not have any event horizon covering the singularity. On the contrary, the singularity itself is a null singular point. Therefore, the singularity is visible to the outside observer at future timelike infinity~($I^{+}$).
\begin{figure}[hbt!]
{\includegraphics[width=85mm]{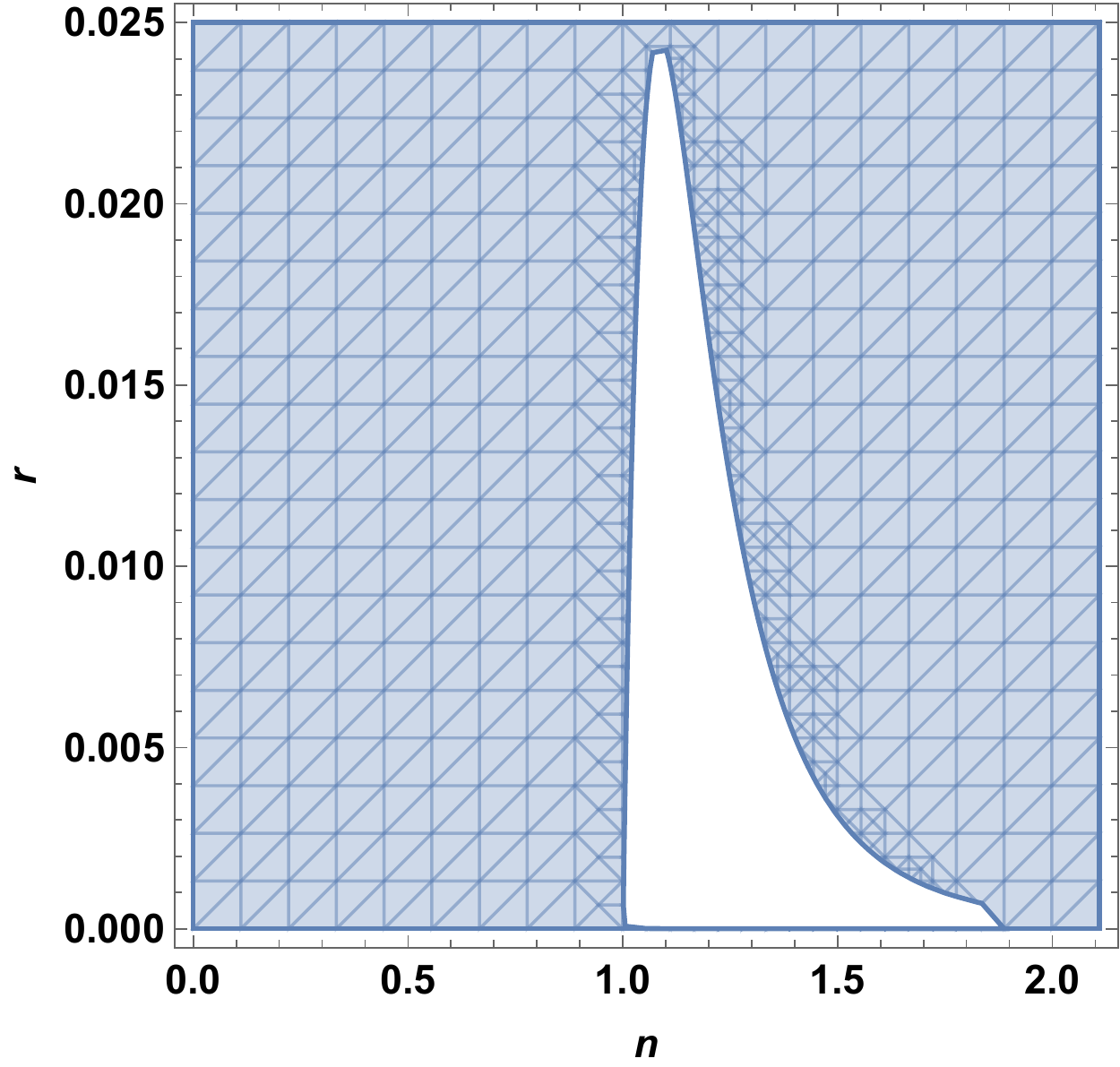}}
\caption{Figure shows range of weak energy condition violation(i.e. white region). Therefore, the condition $\rho+P_r>0$ is satisfied for any value of $r$ when $ n \leq 1 $.  }
\label{fig:regionplot}
\end{figure}

Nevertheless, in order to be a physically valid solution of Einstein's field equations, the dark matter spacetime should satisfy the weak energy conditions. Using Einstein's field equations, we can compute the energy density and pressures of dark matter spacetime as:
\begin{equation}
\rho = \frac{\alpha(1+\alpha) + \beta (1-n+2\alpha)(2- \alpha) + \beta^{2} (2-\alpha)^2}{\kappa r^2 \left(1 + \alpha + \beta (2-\alpha) \right)^2}\, , \label{rho}
\end{equation}
\begin{equation}
P_r = 0 \label{Pr}\,,
\end{equation}
\begin{equation}
P_\theta = P_\phi = \frac{1}{4}\left(\alpha + \beta (2-\alpha)\right) \rho\,\, . \label{Ptheta}
\end{equation}

 In Fig.~(\ref{fig:regionplot}), the permitted region (i.e., the shaded region) for which weak energy conditions are fulfilled is depicted in the space of $n$ and $r$. One can observe that the weak energy conditions are satisfied for $n \leq 1$ and violated for $n>1$. Therefore, throughout the paper, we consider the value of parameter $n=1$. 
\begin{figure}
{\includegraphics[width=85mm]{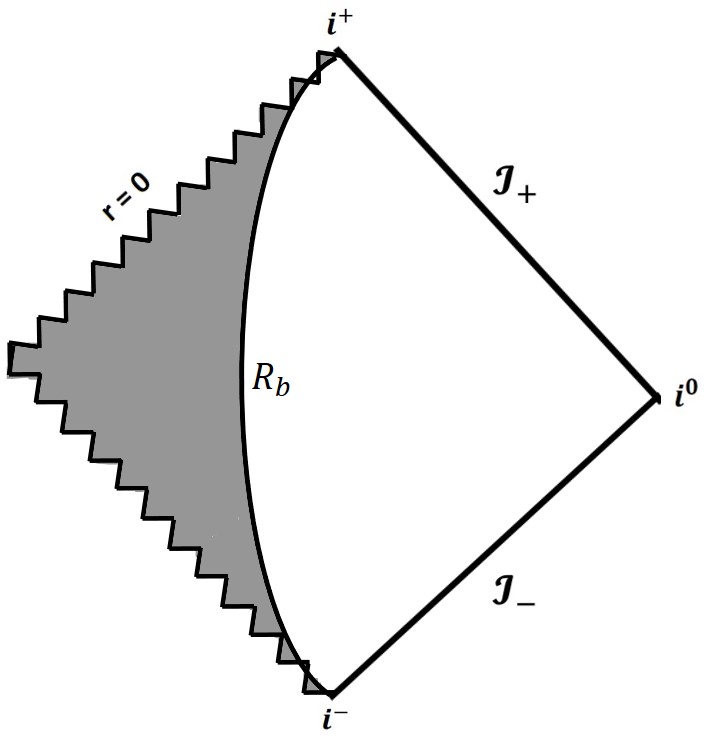}}
\caption{Carter-Penrose diagram of the dark matter spacetime written in Eq.~(\ref{dm_metric}) for $n=1$. The nulllike singularity at $r=0$ is shown by black zigzag lines. The corresponding future null infinity ($\mathcal{J}_{+}$), past null infinity ($\mathcal{J}_{-}$), past timelike infinity ($i^{-}$), future timelike infinity ($i^{+}$), and spacelike infinity ($i^{0}$) are shown. Here $R_{b}$ is matching radius with constant radial distance. Gray colour show the interior dark matter spacetime.}
\label{pd_dm}
\end{figure}

 The causal nature of the singularity can be ascertained by studying the Penrose diagram of spacetime. In a Penrose diagram, the temporal and radial coordinates of the metric are transformed in such a way that the entire spacetime manifold can be adequately represented in a finite-sized causal diagram. In Fig.~(\ref{pd_dm}), we show the Penrose diagram of the dark matter spacetime for $n=1$. From the figure, it can be concluded that the singularity at $r=0$ in spacetime~(\ref{dm_metric}) is nulllike in nature. 

Next section onwards, we shift our attention to the observational aspects of our proposed dark matter spacetime with a nulllike singularity at the galactic center.

\section{Observational constraints on Dark matter spacetime from the orbital dynamics of S2 star}
\label{section3}
Before going into the orbital dynamics of S2 star and the corresponding observational constraints on dark matter spacetime, in the next subsection, we derive the orbit equation of a test particle freely falling in a spherically symmetric and static spacetime (Eq.~(\ref{metric})).
\subsection{Timelike geodesics in dark matter spacetime}
In this subsection, we study the timelike bound orbit of the test particle in the static and spherically symmetric spacetime given by Eq.~(\ref{dm_metric}).

 For bound orbits, the total energy of the particle is greater than or equal to the minimum of the effective potential, i.e., $E\geq V_{min}$. For the minimum of the effective potential, following condition should be obeyed
\begin{equation}
    \frac{d V_{eff}}{dr}=0\,,\,\,\,\,\,\frac{d^2 V_{eff}}{dr^2}\bigg|_{r_s}>0\,,
\end{equation}
where $r=r_s$ is the radius at which $V_{eff}(r)$ has a minimum value and the stable circular orbit of a massive particle exists. Here, the $V_{eff}(r)$ is given by Eq.~(\ref{vr}). The bound elliptical orbits of the particle exist for $V_{eff}<E<0$. For the bound orbit, the periapsis $(r_{min})$ and apoapsis $(r_{max})$ of the orbit can be calculated using the condition $ V_{eff} - E = 0$. Therefore, one can define the bound orbits of the freely falling particles in the following way,
\begin{eqnarray}
   V_{eff}(r_{min})=V_{eff}(r_{max})=E\, , \,\, \nonumber\\
   E-V_{eff}(r)>0\, ,\,\,\,\forall r\in (r_{min},r_{max}).
   \label{bound}
\end{eqnarray}

For some given conserved values of $h$ and $E$, the shape of the timelike orbit can be determined by describing how the radial coordinate $r$ changes with the azimuthal coordinate $\phi$. Using Eq.~(\ref{E}) and the metric in Eq.~(\ref{metric}), we can write
\begin{equation}
    \frac{d\phi}{d r}=\frac{h}{r^2}\frac{\sqrt{g_{rr}(r) g_{tt}(r)}}{\sqrt{2 (E-V_{eff})}}\,.
\end{equation}
Using the above equation, the orbit equation for a static and spherically symmetric spacetime can be written as
\begin{widetext}
 \begin{equation}
  \frac{d^2u}{d\phi^2}+\frac{u}{g_{rr}(u)}-\left[\frac{u^2}{2g_{rr}^2(u)}+\frac{c^2}{2g_{rr}^2(u)h^2}-\frac{e^2}{2g_{tt}(u)g_{rr}^2(u)h^2}\right] \frac{dg_{rr}(u)}{du}+\frac{e^2}{2g_{tt}^2(u)g_{rr}(u)h^2}\frac{dg_{tt}(u)}{du}  = 0\,\, ,
  \label{orbiteqgen}
    \end{equation}
\end{widetext}
where $u(\phi)=1/r$. We solve the orbit equation numerically to describe the shape of the bound orbit of the test particle in the dark matter spacetime. 

%

\subsection{Orbital dynamics:} To trace the orbital path of the star, first, we have to solve the time-varying geodesic equation, which gives the equations of motion of the test particle - 
\begin{equation}
  \dot{t} = \sqrt{\frac{E}{c^2 g_{tt}^2 r(\tau)}}, \label{tdot}
\end{equation}
\begin{equation}
    \ddot{r} = \frac{1}{2 g_{rr}(r(\tau))}\left[- c^2 \frac{d g_{tt}( r(\tau))}{dr}  \dot{t}^2 - \frac{d g_{rr} (r(\tau))}{dr}
 \dot{r}^2 + 2r\dot{\phi}^2\right],
 \label{rdotdot}
\end{equation}
\begin{equation}
    \dot{\phi} =  \frac{h}{r(\tau)^2},\label{phidot}
\end{equation}
The numerical solutions of Eqs.~(\ref{rdotdot}) provide the orbital positions over the time.
\begin{figure}
\includegraphics[width=\hsize,clip]{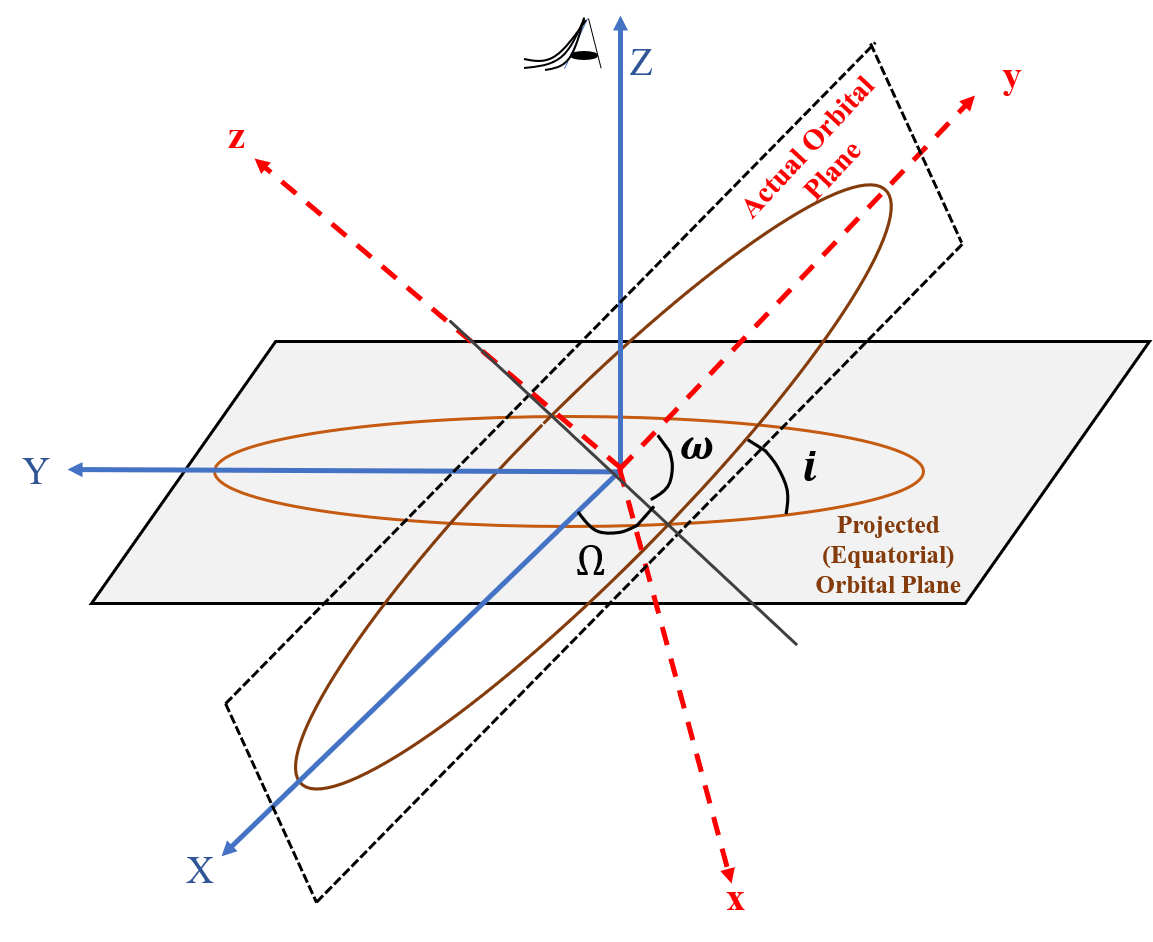}

\caption{This figure shows a projection of the actual orbital plane onto the projected orbital plane seen by the asymptotic observer. The focus of the ellipse gives the location of the Sgr A* compact object. Here, the Z-axis of the coordinate system is defined by the vector pointing from the galactic center to the solar system.}\label{fig:orbit_planes}
\end{figure}
\begin{figure*}[hbt!]
\centering
\subfigure[]
{\includegraphics[width=75mm]{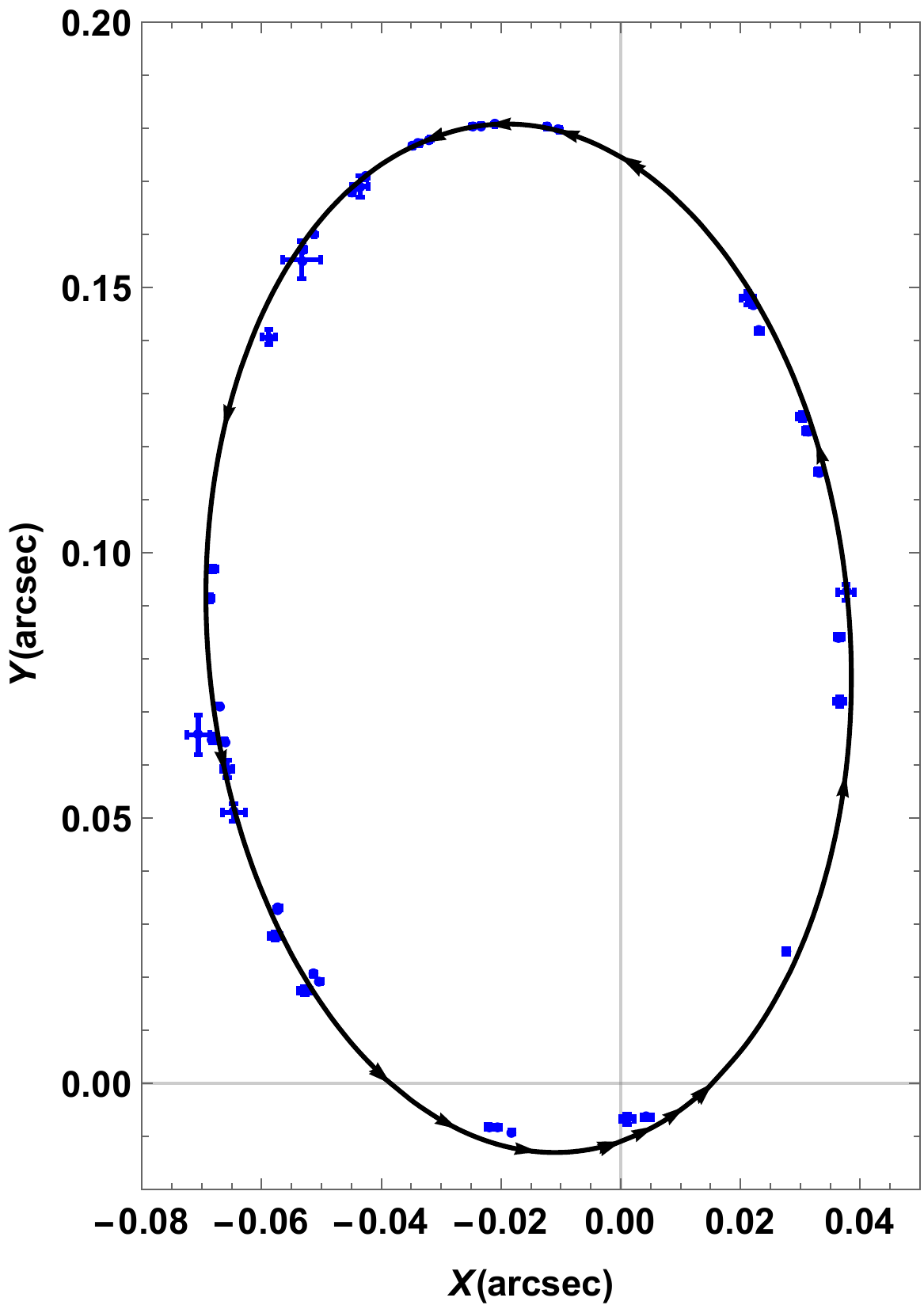}\label{fig5a}}
\hspace{1cm}
\subfigure[]
{\includegraphics[width=85mm]{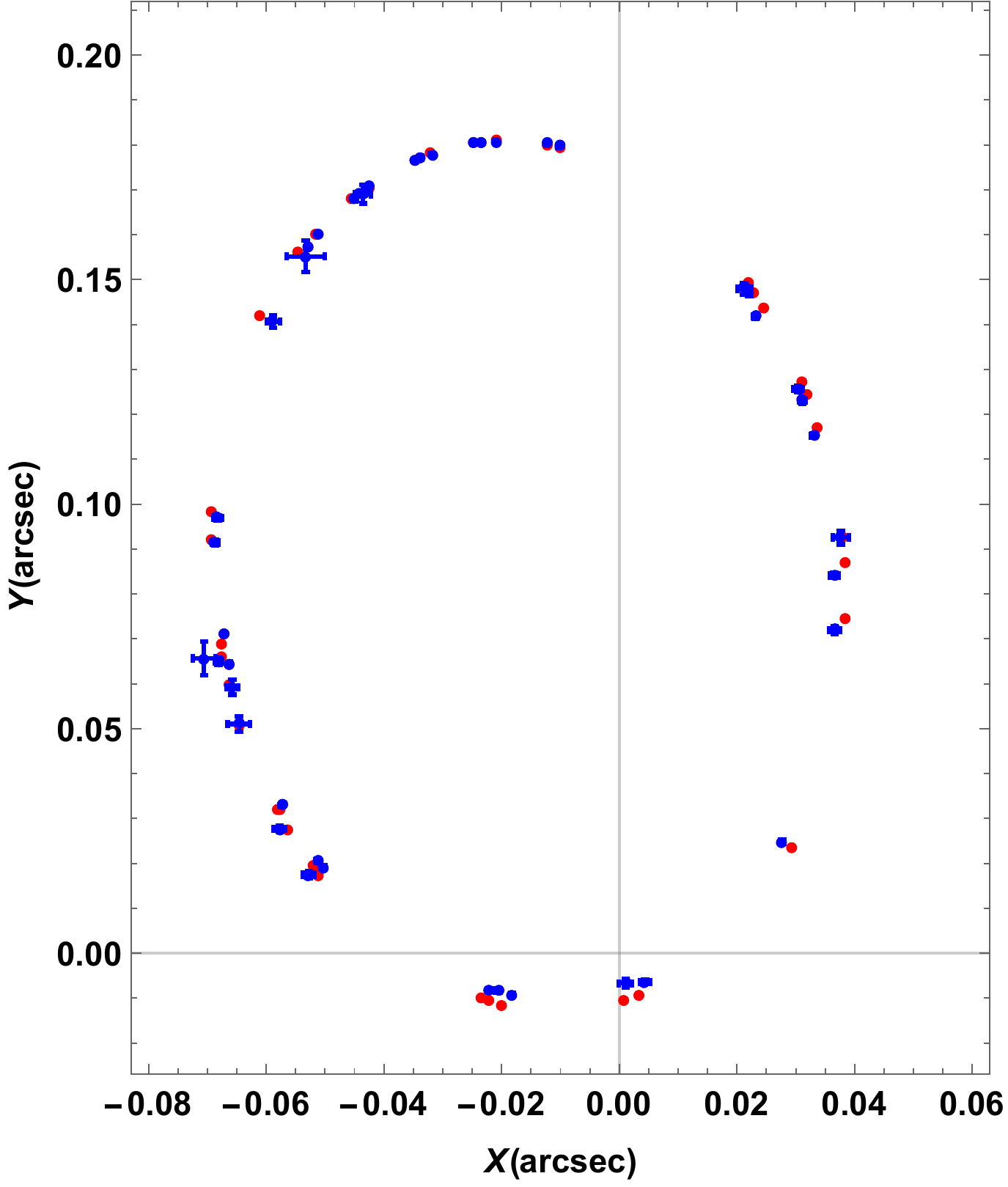}\label{fig5b}}
 \caption{Theoretical and observed orbits of the S2 star around SgrA*. In Fig.~\ref{fig5a}, blue crosses correspond to the original data of the S2 star, and the black curve represents the best fit dark matter spacetime timelike orbit with the orbital data. Fig.~\ref{fig5b} represents the comparison between the experimental data points and the theoretically fitted data points. Here, the theoretical orbit is calculated by solving the equations of motion of a test particle in dark matter spacetime. We use the observational data given in~\cite{datasupplementary}}
\label{fig5}
\end{figure*}
\begin{figure*}[hbt!]
\centering
\subfigure[]
{\includegraphics[width=75mm]{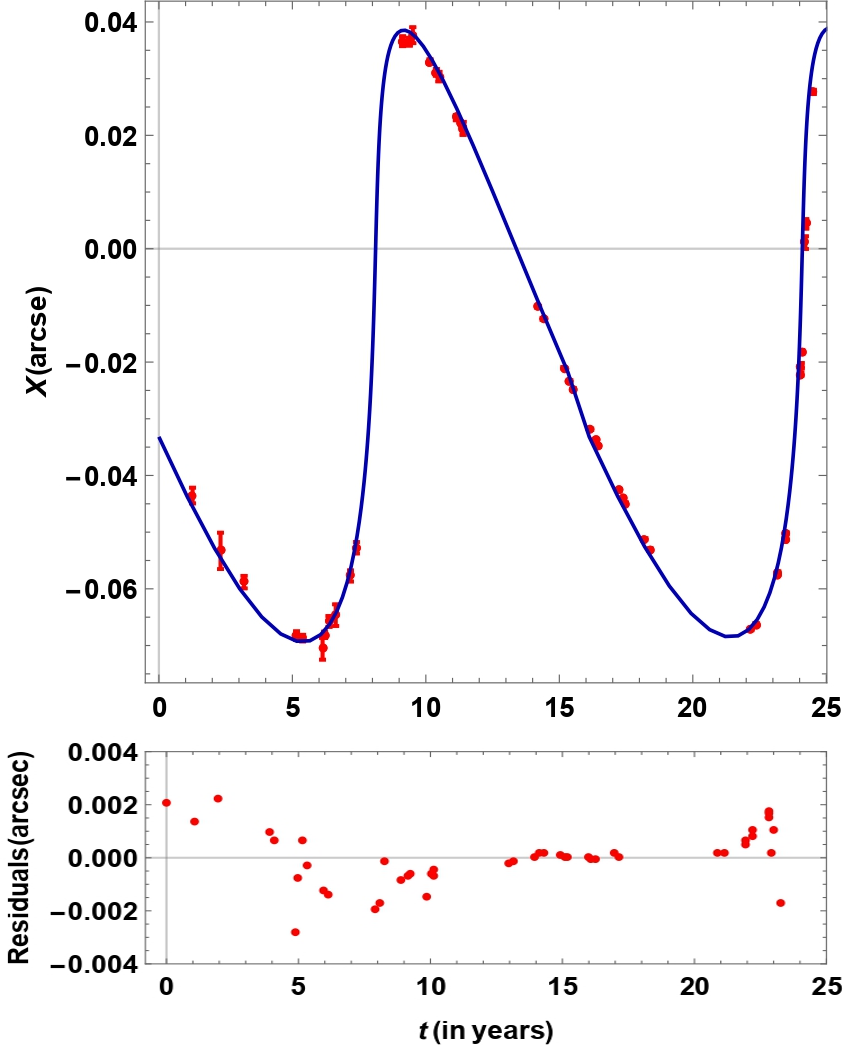}\label{fig6a}}
\hspace{1cm}
\subfigure[]
{\includegraphics[width=75mm]{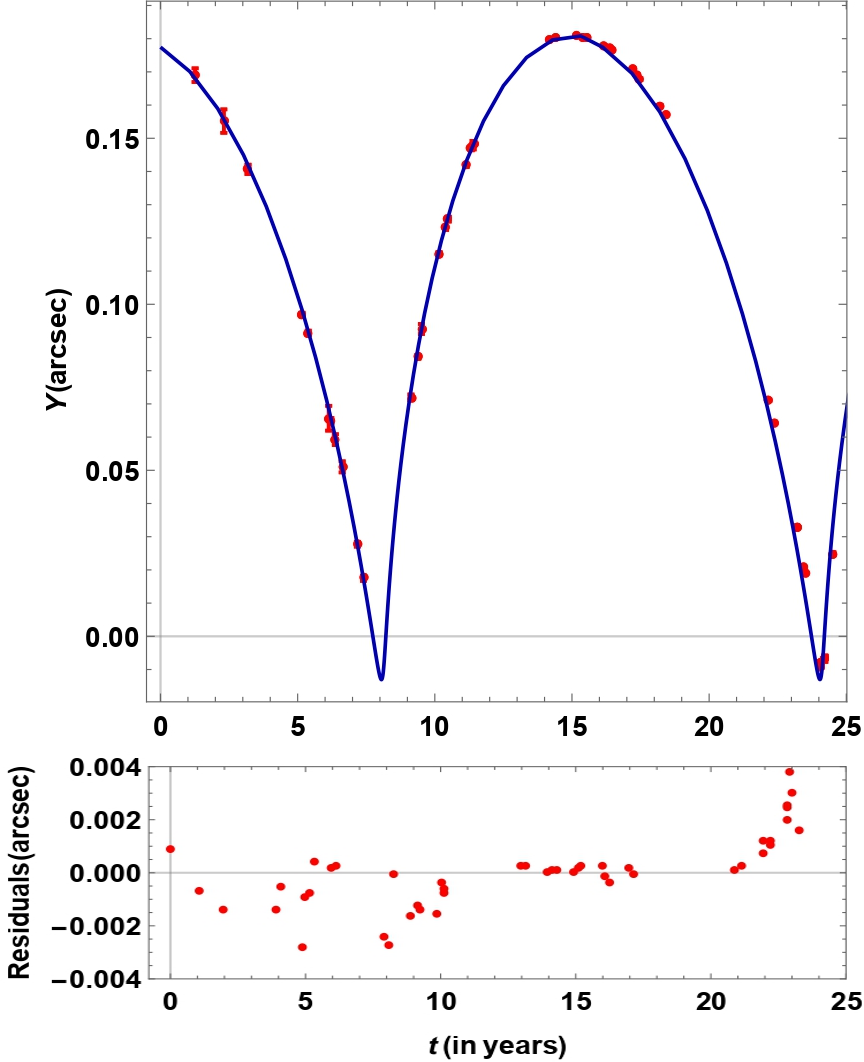}\label{fig6b}}
 \caption{Plot of $X$ and $Y$ as a function of time, along with the respective residuals of the best fit for the dark matter spacetime. In Figs.~\ref{fig6a} and~\ref{fig6b}, red crosses correspond to the original data of the S2 star, and the blue curves represent the best fit of the theoretical prediction of dark matter spacetime.}\label{fig6}
\end{figure*}
\begin{table*}
    \centering
    \caption{\textbf{Summary of best-fit values of the model and the orbital parameters for S2 within the Dark matter(DM) model}}
    \begin{tabular}{l c c}
        \hline\hline
Parameter   &   Dark Matter spacetime   \\
     &       \\
\hline
Total mass of Dark matter, $M\odot$  &    $3.947*10^{12}$ \\
Distance to pericenter, $(arcsec)$  &    $0.000552$   \\
Distance to apocenter, $(arcsec)$  &    $0.009665$   \\
eccentricity,   &    $0.89194$   \\
\hline
$h$, $pc(km/s)$  &    $4.4499392$   \\
$E$, $(km/s)^{2}$  &    $8.99956211840861*10^{10}$   \\
$R_{b}$, $parsec(pc)$  &    $976693$  \\
$V_{b}$, $km/s$  &    $131.826$  \\
$r_{ph}$, $pc$  &    $2.10256*10^{-7}$   \\
Argument of Pericenter, $\omega$~($^{\circ}$)  &    $133.902$   \\
Inclination, $i$~($^{\circ}$)  &     $229.935$  \\
Ascending Node, $\Omega$~($^{\circ}$)  &    $68.6008$ \\
$R_{0}$, $pc$  &    $8196.4033$  \\
Initial time, $yr$  &    $1.2619854$  \\
\hline
Orbital Period, $P$~($yr$)  &    $16.2553$  \\
Shadow Size, ($pc$)  &    $5.71538*10^{-7}$   \\
Precession angle, ($arcminutes$)  &    $21.6'$   \\
\hline
{$\bar{\chi}^2_X$} &    $3.92665$  \\
{$\bar{\chi}^2_Y$} &    $7.91589$   \\
{$\langle \bar{\chi}^2\rangle$} &    $5.92127$   \\

\hline
    \end{tabular}
    \label{tabl1}
\end{table*}
\vspace{1cm}

In terms of the Cartesian coordinates, we express the position and velocity of a particle moving along a real orbit as $(x, y, z)$ and $(v_x, v_y, v_z)$ respectively. For $\theta=\pi/2$ case, these can be obtained by transforming the coordinates from spherical Schwarzschild to cartesian:
\begin{eqnarray}
    x = r \cos\phi,\,\,\,\,\,\,\,
    y = r \sin\phi,\,\,\,\,\,\,\,
    z = 0\,.
\end{eqnarray}
%
The solution of the orbit equation~(Eq.\,(\ref{rdotdot})) allows us to trace the stellar position, but in order to compare it to observational data, we must first establish the relation between the real orbit and the apparent orbit in the observer's sky. 
We use classic Thiele-Innes constants, that are $A, B, C, F, G$, and $H$, to obtain the theoretical apparent orbit on the plane of sky given by coordinates ($X, Y, Z$) from the real orbit coordinates ($x, y, z$)~(see Fig.~(\ref{fig:orbit_planes})).
\begin{eqnarray}
    X &=&  x\,B + y\, G,\\
    Y &=& x\, A + y\, F,\\
    Z &=& x\, C + y\, H,
\end{eqnarray}
where,
\begin{eqnarray}
    A &=& \cos\Omega \cos\omega - \sin\Omega \sin\omega \cos i,\\
    B &=& \sin\Omega \cos\omega + \cos\Omega \sin\omega \cos i,\\
    C &=& \sin\omega \sin i,\\
    F &=& -\cos\Omega \sin\omega - \sin\Omega \cos\omega \cos i,\\
    G &=& -\sin\Omega \sin\omega + \cos\Omega \cos\omega \cos i,\\
    H &=& \cos\omega \sin i,
\end{eqnarray}
where, $\omega$, $i$, and $\Omega$ are the argument of the pericenter, the inclination between the real orbit and the observation plane, and the ascending node angle, respectively. 

General Relativity predicts that a star orbiting close to a supermassive compact object is supposed to experience a relativistic redshift. A total redshift~($z(r)$) is a combination of the special relativistic Doppler shift and the gravitational redshift. The redshift function can be written as~\cite{Becerra-Vergara:2020xoj}
\begin{equation}
    z = \gamma - 1 + u_{Z}, \label{redshift}
\end{equation}
where,
\begin{equation}
    u_{Z} = (\dot{r} \sin(\phi + \omega) + r \dot{\phi}\cos(\phi + \omega))\sin i
\end{equation}
is the apparent four-velocity of the emitter, and
$\gamma=\frac{dt}{d\tau}=\frac{\sqrt{E}}{c g_{tt}}$ is the Lorentz factor. In the weak field limit, $\gamma \to 1 $. Therefore, the redshift is given by the Keplerian (Newtonian) contribution ($z_{K}$), where
\begin{equation}
   z_K \equiv u_Z\,.
\end{equation}

Here we consider the gravitational constant~($G$) and the velocity of light~($c$) as $4.30091\times10^{-3} pc\, M_\odot (km/s)^2$ and $3\times10^{8} km/s$ respectively. Using these universal constants, we get the radial distance in the parsec unit.

\subsection{Orbit of the S2 star around the Sgr A*}
The motions of the S-stars that comprise the nuclear cluster at the centre of the Milky Way have been closely monitored for over three decades. Their trajectories revealed the presence of a supermassive compact object with a mass of ~$4 \times 10^6 M_{\odot}$ at the cluster centre. The most important S-cluster member is S2, which has an orbital period of about 16 years around Sgr A*.
It made its closest approach in May 2018, at a distance of 120 astronomical units (au) from the Sgr A* with a velocity of $2.7\%$ of the speed of light.
This close proximity of S2 to the Sgr A* compact object causes the relativistic redshift, which is the combination of the transverse Doppler shift from special relativity and the gravitational redshift from General relativity. Here, we use the astrometric data of the S2 star, which were obtained using speckle imaging with NIRC\,(Near-Infrared Camera) on Keck I~(1995–2005) and Adaptive Optics~(AO) imaging with NIRC2\,(Near-Infrared Camera 2) on Keck II (2005-2018)~\cite{datasupplementary}. In this paper, we use the data of the astrometric positions of S2 given in the supplementary material of \cite{datasupplementary}. Here, we want to determine the values of the free parameters of the theory from the astrometric data of the S2 star.   
\begin{figure}[ht!]
{\includegraphics[width=85mm]{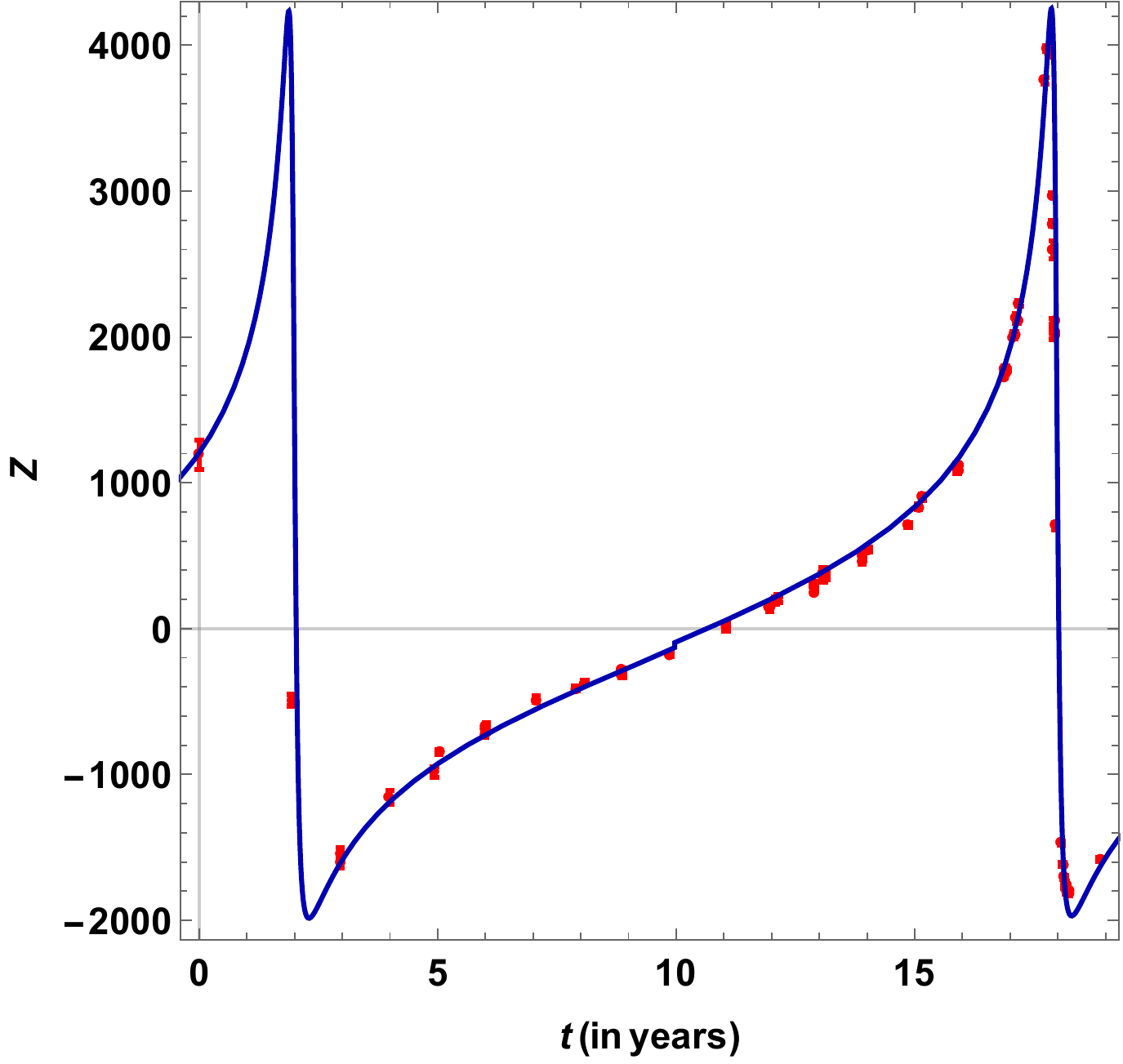}}
\caption{Figure shows the fitting of red shift data with dark matter spacetime. The blue line represents theoretical fitting, while the red dot represents experimental data. The observational data is taken from~\cite{datasupplementary}.}
\label{fig7}
\end{figure}
\subsection{Parameter estimation from Orbital dynamics}
In the present paper, we have used Metropolis Hastings Monte Carlo Markov Chain Algorithm (MH-MCMC) to find the best fit values of parameters to the astrometric data of S2 star \cite{mcmc1,mcmc2}. Let us consider a parameter $\psi$ which is random with normal distribution $\pi(\psi)$. where $\pi(\psi)$ is prior distribution which gives us prior uncertainty regarding $\psi$. To estimate best fit parameter using the posterior analysis with some data ($\mathcal{D}$), that is, using the Bayesian inference 
\begin{equation}
\pi(\psi | \mathcal{D}) = \frac{\pi(\psi) p(\mathcal{D} | \psi)}{p(\mathcal{D})} \propto \pi(\psi) p(\mathcal{D} | \psi) \,,
\end{equation}
where $\pi(\psi)$ is the prior, $p(\mathcal{D} | \psi)$ is the likelihood and $p(\mathcal{D})$ is marginal probability density function (pdf) of $\mathcal{D}$ -
\begin{equation}
    p(\mathcal{D}) = \int_{\psi}^{} \pi(\psi) p(\mathcal{D} | \psi)  \,d\psi \,,
\end{equation}
which can be regarded as a normalizing constant as it is independent of $\psi$. The posterior pdf is thus proportional to the product of prior and likelihood.

\begin{figure*}[ht!]
\centering
\subfigure[]
{\includegraphics[width=82mm]{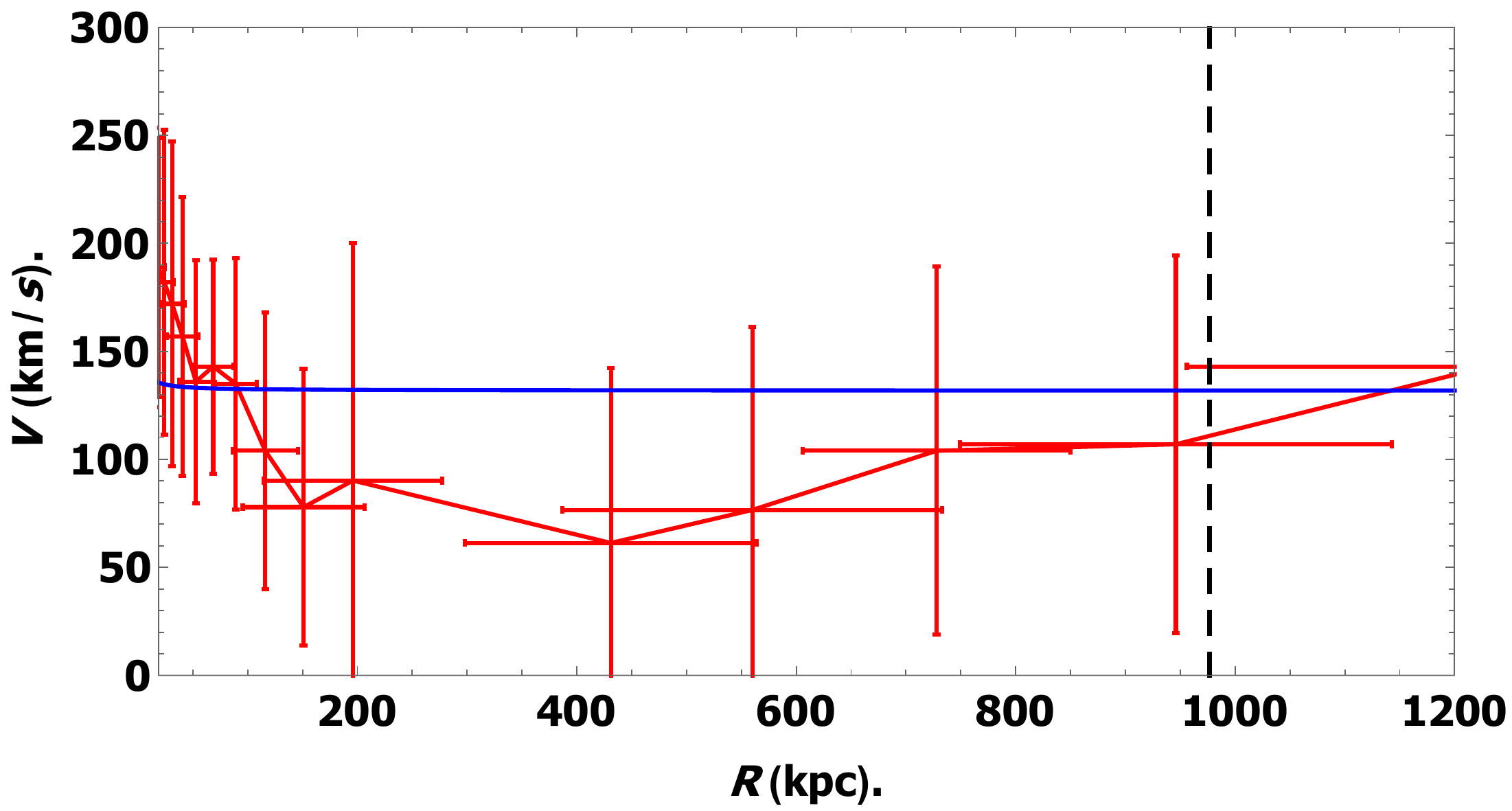}\label{fig3a}}
\hspace{1cm}
\subfigure[]
{\includegraphics[width=82mm]{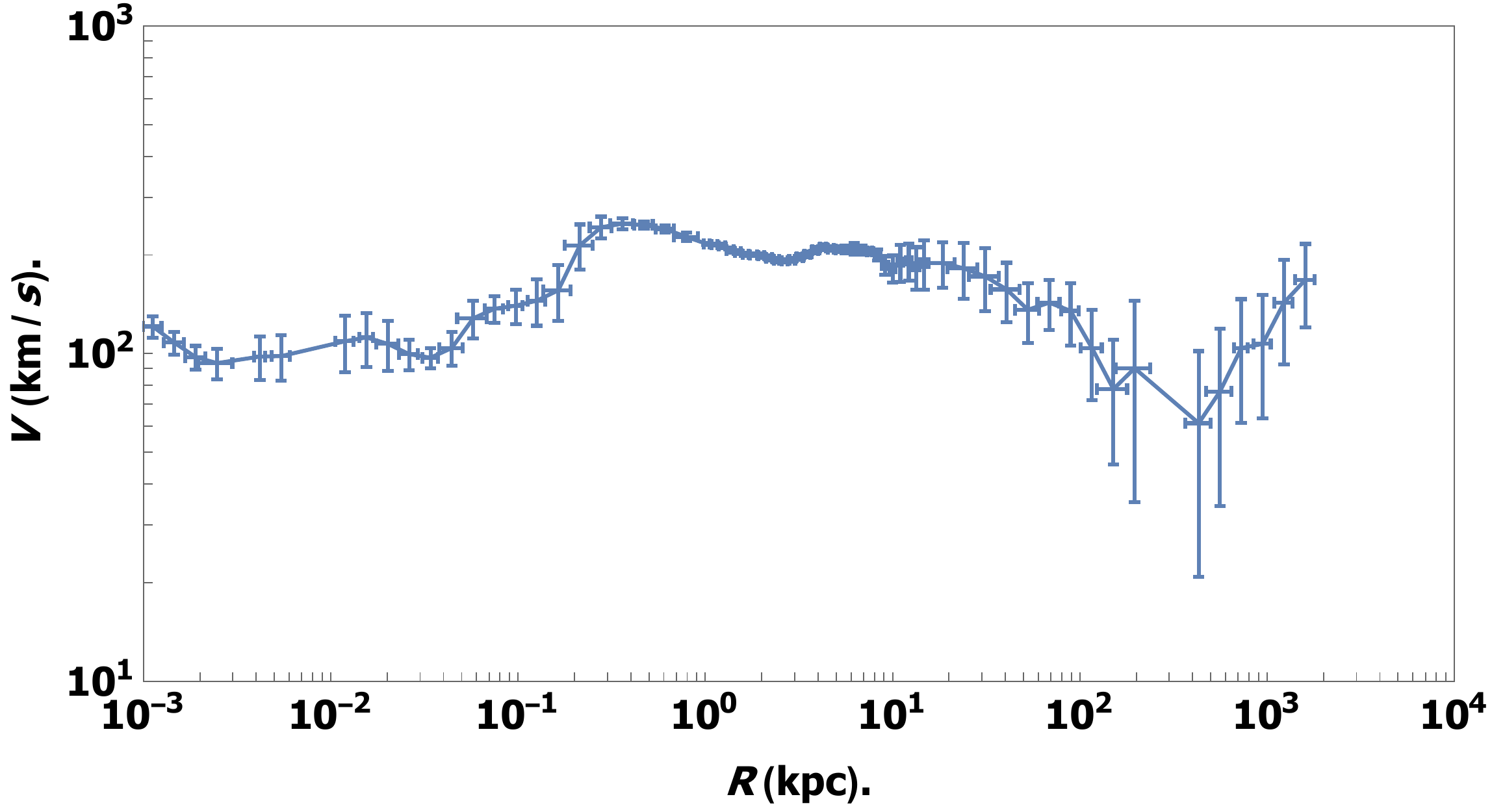}\label{fig3b}}
 \caption{Rotation curve of the Milky Way galaxy. The red crosses in Fig.~(\ref{fig3a}) correspond to the original data of a flat velocity curve, and the blue line represents the best fit velocity curve in the range of $20 kpc$ to $976.6 kpc$. The black dotted line represents the estimated boundary of the halo radius. Fig.~(\ref{fig3b}) represents the logarithmic plot between the experimental data of the velocity of stars with radial distance.}\label{fig3}
\end{figure*}

\subsubsection{priors}
In the model, we considered 10 free parameters. The prior distribution~($\pi(\psi)$) is normally distributed over parameter $\psi$ with an initial value $\mu$ and dispersion $\sigma$, $\mathcal{N(\mu,\sigma)}$. Here, we adopt six informative initial values of parameters from the previous paper~\cite{Becerra-Vergara:2020xoj} and four non-informative initial values. The distance from earth to the galactic center ($R_{0}$), argument of the pericenter ($\omega$), inclination ($i$) and ascending node ($\Omega$) are $8190 pc$, $66.4^{\circ}$, $134.3^{\circ}$ and $227.9^{\circ}$ respectively which are directly taken from the paper \cite{Becerra-Vergara:2020xoj}. While using the semi-major axis ($a$)(0.1252 arcsec) and eccentricity($\epsilon$)(0.88), we have calculated the distance to the pericenter and apocenter, using $r_{min} = a(1-\epsilon)$ and $r_{max} = a(1+\epsilon)$, respectively.

For the two known turning points of the orbit, which are $r_{min}$ and $r_{max}$, we can obtain angular momentum~($h_c$) and energy~($E_c$) at the turning points using Eq.~(\ref{bound})
\begin{equation}
    h_{c} = \sqrt{\frac{c^2 r_{min}^2 r_{max}^2 (g_{tt}(r_{min})-g_{tt}(r_{max}))}{(r_{min}^2 \, g_{tt}(r_{max}) - r_{max}^2 \, g_{tt}(r_{min}))}}\,,\label{hc}
\end{equation}
\begin{equation}
   E_{c} = g_{tt}(r_{min}) \left(c^2 + \frac{h_{c}^2}{r_{min}^2} \right).\label{Ec}
\end{equation}

For non-informative initial values, that is, matching radius ($R_{b}$), flat velocity at matching radius ($V_{b}$), radius of photon sphere ($r_{ph}$) and initial time, we consider intuitive values with previous observation, which are $8190pc$, $150km/s$, $3GM/c^{2}$ and $1$ year, respectively.

Using the above values, we can numerically integrate the equations of motion, Eqs.~(\ref{tdot})-(\ref{phidot}) giving appropriate initial conditions at the initial proper time. Once the initial conditions are established, the coordinates of the orbit of S2 at any time $t$ can be predicted.

%
\subsubsection{Likelihood}

In order to calculate the goodness of fit, we compute the reduced $\chi^2$ value for each of the observables. The likelihood function ($p(\mathcal{D} | \psi)$) is used in the MH-MCMC analysis with some symmetric error

\begin{equation}
-\log \mathcal{L} \propto \sum_i \left[\left(\frac{X_{e,i} - {X}_{i}}{\sigma^X_i}\right)^2 + \left(\frac{Y_{e,i} - {Y}_{i}}{\sigma^Y_i}\right)^2 \right].
\end{equation}
where, ($X_{e}, Y_{e}$) are the observed astrometric data and ($\rho_{X}, \rho_{Y}$) are the error values in the observed data. $X$ and $Y$ are theoretical calculated values of trajectories of S2 star.

\subsubsection{$\chi^{2}$ analysis}
We do not sample the posterior distribution of the model, but rather a minimum $\chi^2$ value. For numerical $\chi^2$ computation, we have taken 3 lakh data points from the normal distribution to calculate the minimum $\chi^2$ value. We ran six chains for 50000 steps and discovered that the stationarity of data converged to the lowest $\chi^2$ value around 5.92 for the astrometric data of the S2 star. In our computational chain, initial $\chi_{i}^2$ is grater than the final $\chi_{f}^2$. Therefore, if $e^{(\chi_{i}^2 - \chi_{f}^2)} > 1$, we accept the jump, if it not, it is rejected.

%
From the observer plane, we fit ($X, Y$)( see in Fig.~(\ref{fig6a}) and Fig.~(\ref{fig6b}) with residuals). We also fit the redshift factor $Z(r)$, as shown in the Fig.~(\ref{fig7}). Here, the total $\chi^2$ value can be taken as the average of the two: 
\begin{equation}
    \langle\chi^2\rangle \equiv \frac{1}{2}\left( \bar{\chi}^2_{X} + \bar{\chi}^2_{Y} \right).
\end{equation}

All the 10 fitted parameter values are tabulated in Table~\ref{tabl1}. By using those parameters, we calculate the mass, eccentricity, and distance to the pericenter and apocenter. We have also investigated the precession angle, which is $21.6$ arcminutes per orbital period. The best-fit value of $r_{ph}$ provides information about the shadow size. As we have now determined the best-fit values of the parameters for the Milky Way galaxy, we move forward to analyse the redshift, rotation profile, and shadow property of the dark matter spacetime using those best-fit values.

Fig.~(\ref{fig7}) shows the redshift function $z$ for the dark matter model for the S2 star orbit. The theoretical redshift function $z$ is computed using the Eq.~(\ref{redshift}) by taking into account the best-fit values of parameters shown in Table~(\ref{tabl1}). 
\section{Galactic rotation curve}
\label{section4}
The rotational velocity of the particle in the dark matter spacetime can be obtained by substituting $g_{tt}(r)$ from Eq.~(\ref{gtt}) in Eq.~(\ref{velocity}):
\begin{equation}
    V_{c} = \sqrt{\frac{V_{b}^2 (r - r_{ph}) + c^2 r_{ph}}{r}} \,,\label{Vc}
\end{equation}
where, $V_{b} = \sqrt{\frac{G M}{R_b}}$. Here, $G$, $R_b$, and $M$ are the gravitational constant, halo radius, and total mass of the galaxy enclosed in the halo radius, respectively. Here, the circular velocity profile changes with the radial distance. From Eq.~(\ref{Vc}), one can see that the circular velocity $V_{c} = V_{b}$ near $r=R_b$ where $R_b>>r_{ph}$. 
For the velocity profile of the Milky Way galaxy, here we use the data provided in the paper~\cite{Sofue:2013kja}. Fig.~(\ref{fig3a}) shows the fitting of the theoretical rotation curve with the observations of the Milky Way galaxy. In Fig~(\ref{fig3a}), the red crosses correspond to the rotational velocity data of our Milky Way galaxy~\cite{Sofue:2013kja}, and the blue line shows the theoretical circular velocity of the test particle calculated from Eq.~(\ref{Vc}) using the best-fit parameters. The black dotted vertical line at $r=976kpc$ represents the boundary of DM spacetime. We find that the fitting is fairly good, with observations in the region of $20kpc$ to $976kpc$. Therefore, this model of dark matter satisfies the flat rotation curve of the Milky Way far away from the galactic center. The full range of data from $1pc$ to $1000kpc$ on the logarithmic scale is represented in Fig.~(\ref{fig3b}). Our model is unable to explain the rotation curve in the region of $0$ to $0.3kpc$. This is due to our consideration that dark matter dominates over the large distances around the halo radius.
\begin{figure*}
\centering
\subfigure[Intensity distribution in dark matter spacetime.]
{\includegraphics[width=82mm]{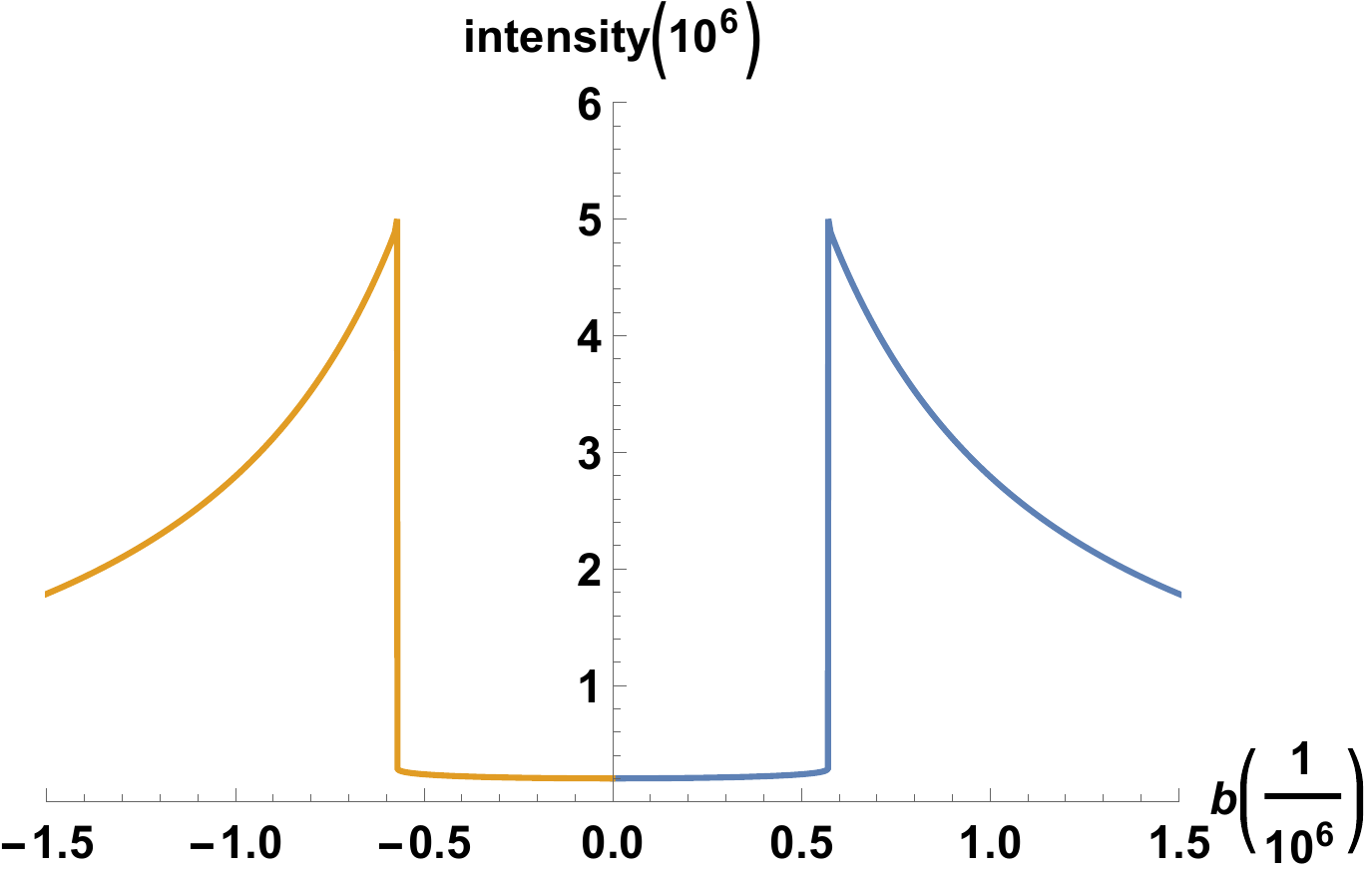}\label{fig83}}
\hspace{0.2cm}
\subfigure[Shadow in dark matter spacetime.]
{\includegraphics[width=68mm]{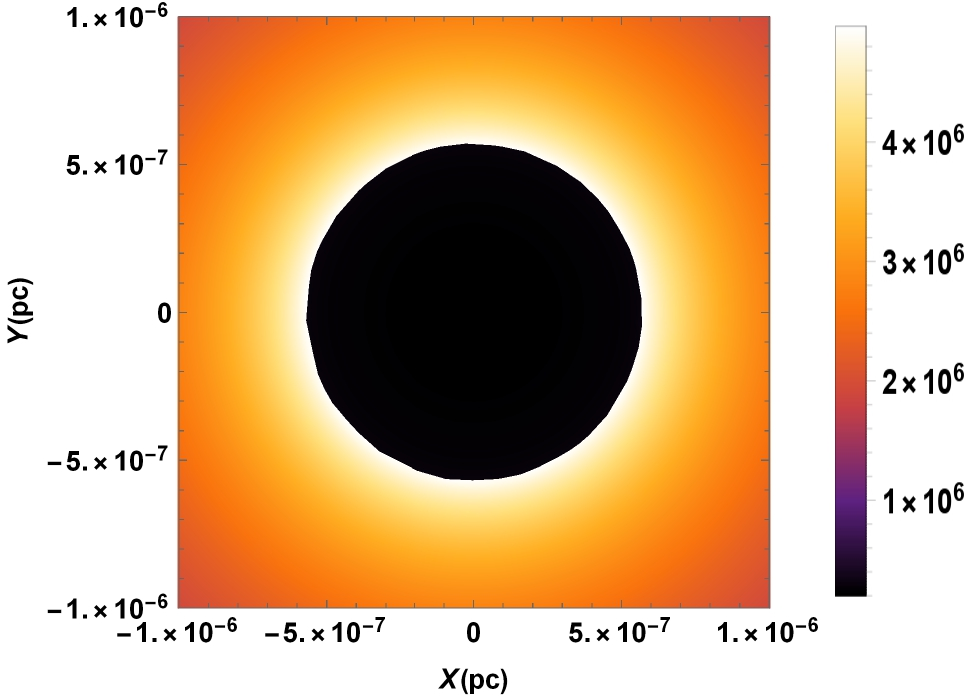}\label{fig84}}
 \caption{In this figure, the intensity map in the observer's sky (left column figure) and the shadow of the central object (right column figure) are shown for dark matter spacetime. The radius of the shadow of dark matter spacetime (Fig.~(\ref{fig84})) is $5.7154*10^{-7} $pc}
\label{fig8}
\end{figure*}
\section{Shadow of the proposed dark matter spacetime}
\label{section5}
The shadow of any spacetime geometry depends upon the nature of the effective potential of the null geodesics in that spacetime. As mentioned in Section~(\ref{section2}), the effective potential for the null geodesics in the equatorial plane is given by Eq.~(\ref{veff}). At the turning point of the photon, where $\dot{r}=0$ and $V_{eff}(r_{tp})=\frac{1}{b_{tp}^2}$, we can write the impact parameter~($b_{tp}$) corresponding to the turning point as,
\begin{equation}
    b_{tp}=\frac{r_{tp}}{\sqrt{g_{tt}(r_{tp})}}\,.
\end{equation}

When there exists a maximum value of the effective potential of lightlike geodesics in spacetime, and that maximum value is also the upper-bound of the potential, the minimum impact parameter of a photon is the impact parameter associated with the photon sphere ($b_{ph}$). This minimum impact parameter is the critical impact parameter ($b_{crit}=b_{ph}$), which differentiates photons colliding with the central object from those reaching a minimum distance and returning to infinity. Photons from a remote source with an impact parameter b greater than the critical impact parameter $b_{ph}$ are scattered and reach the  observer. On the other hand, photons with impact parameters less than the critical impact parameter are trapped inside the photon sphere and never reach the observer, resulting in a dark area or shadow in the observer's sky. Therefore, the shadow in the observer's sky appears to be a circular 2-dimensional dark disc with a radius of $b_{crit}$.

As we know, the shadow of a massive object forms due to the accreting matter surrounding it. Therefore, we need to compute the intensity of light emitted by the accreting matter to obtain the shadow of the massive object in the observer's sky. In order to get the intensity map of the emitting area, we need to consider various radiating processes and emission mechanisms. The measured intensity at the observer's sky point (X, Y) can be given as~\cite{Bambi:2013nla},
\begin{equation}
    I_{\nu_{obs}}(X,Y)= \int_{\gamma}^{} g^3 j(\nu_e) dl_{prop}\,\, ,\label{intensity1}
\end{equation}
where $g=\nu_{obs}/\nu_e$ is the redshift factor, $\nu_{obs}$ is the observed photon frequency, $\nu_e$ is the photon frequency as measured in the rest frame of the accreting gas which is emitting radiation, $j(\nu_e)$ is the emissivity per unit volume in the rest frame of the emitter, and $dl_{prop}=k_\alpha u^\alpha_e d\lambda$ is the infinitesimal proper length in the rest frame of the emitter. The integration is done along the photon path ($\gamma$). The redshift factor can be given by~\cite{Bambi:2013nla}
\begin{equation}
    g=\frac{k_\alpha u^\alpha_{obs}}{k_\beta u^\beta_e}\,,
\end{equation}
where, $u^\alpha_{obs} = (1,0,0,0)$ is the four-velocity of the distant static observer, $u^\beta_e$ is the timelike four-velocity of the emitter, and $\lambda$ is the affine parameter, and $k_{\alpha}$ is the four-velocity of the photon. For simplicity, here we consider a simple model of spherically symmetric accreting gas that is freely falling radially and it is optically thin (i.e. the accreting matter does not absorb photons). In a generic spherically symmetric, static spacetime (\ref{metric}), the components of the four-velocity of a radially freely falling particle can be evaluated as,
\begin{equation}
    u^t_e = \frac{1}{g_{tt}}\,\,, \hspace{5mm} u^r_e = -\sqrt{\frac{(1-g_{tt})}{g_{tt}g_{rr}}}\,\,, \hspace{5mm} u^\theta_e = u^\phi_e = 0 .\label{four}
\end{equation}
Using eq.(\ref{four}), the redshift factor can be written as,
\begin{equation}
    g = \frac{1}{\frac{1}{g_{tt}}-\frac{k_r}{k_t} \sqrt{\frac{(1-g_{tt})}{g_{tt}g_{rr}}}}\,\, ,
\end{equation}
where
\begin{equation}
    \frac{k^r}{k^t} = \sqrt{\frac{g_{tt}}{g_{rr}} \left(1-\frac{g_{tt} b^2}{r^2}\right)}\,\, .
\end{equation}
In the present work, we assume a simple model for the specific emissivity in which the radiation from emitter is monochromatic with the emitter's rest frame frequency $\nu_*$ and radially falls by $1/r^2$
\begin{equation}
    j(\nu_e) \propto \frac{\delta(\nu_e-\nu_*)}{r^2}\,\, ,
\end{equation} 
where $\delta$ is the Dirac delta function.
Now, the eq.~(\ref{intensity}) can be written as \cite{Bambi:2013nla},
\begin{equation}
    I_{obs}(X,Y) \propto - \int_{\gamma}^{} \frac{g^3 k_t dr}{r^2 k^r}\,\, ,
\label{intensity}
\end{equation}
where $I_{obs}(X,Y)$ denotes the intensity distribution in the (X, Y) plane of the observer's sky and $X^2+Y^2=b^2$. Now, using the above equation (Eq.~(\ref{intensity})), we can simulate the shadow.

We employed a technique called `Backward Raytracing' to simulate the shadow, which involves tracing light rays backward in time from the observer to the source.

Using the best-fitting parameters, we construct the shadow of Sgr A* using the dark matter model, which is presented in Fig.~(\ref{fig8}). The radius of the shadow is found to be $5.7154\times 10^{-7} pc$. 

\subsection{Limitation and constraints in the paper}
\begin{itemize}
    \item Here we assume that the S2 star moves in the gravitational potential of a single central compact object with an ADM mass $M$. Therefore, the gravitational effect of the other object inside the S2 is neglected, like the G2 object.

    \item In the paper, we do not consider radio source data. Hence, we are not able to best-fit the observed location of the central radio source for Sgr A*, which may confirm the centre of our Milky Way galaxy. However, the best-fit theoretical prediction of the centroid of mass yields reasonable results that match the existing data. 

    \item Here, data is fitted with only the astrometric position of the S2 star. We have not used spectroscopic data in the fitting procedure.
    
    \item The orbital region of an S2 star is considered in the weak gravitational field in order to use the Thiele-Innes constants. We have neglected the Shapiro time delay, which differs no more than 5 to 6 minutes in the orbital period of the S2 star.  

\end{itemize}
%



\section{Conclusion} \label{section6}
The conclusions from this study can be summarised as follows:
\begin{itemize}
    \item In this paper, we present a new approach for constructing a viable class of spacetimes for galactic dark matter in the framework of general relativity. We show that one can use the concept of the galactic flat velocity curve and the conditions for the existence of a photon sphere to construct the $g_{tt}(r)$ metric coefficient. We match the interior dark matter spacetime with the external Schwarzschild spacetime at the matching boundary $R_b$. We consider the internal fluid has zero radial pressure ($P_r=0$) and non-zero tangential pressures ($P_T$). We show that the proposed spacetime satisfies the weak energy conditions and it has a central null singularity.

    \item Next, we constrain the free parameters of the metric using the data of the astrometric positions of the S2 star around Sgr A*. We obtained the orbit equation, solved it numerically, in order to get particle trajectories in the dark matter spacetime and we compare it with the astrometric data of the S2 star. Using the MCMC algorithm, we obtain the value of photon sphere radius $r_{ph}$ and constant circular velocity $V_b$. The best-fit values derived from the minimum $\chi^2$ fitting show good agreement with the previously known estimated values, such as the galaxy's mass being around $3.947\times10^{12}$ solar mass, the distance from the Earth to the center of the compact object being $R_0= 8196 pc$, and the range of the dark matter halo being between $10kpc$ and $976.7kpc$, etc.

    \item Using the astrophysical best-fit values of the parameters of the proposed dark matter spacetime, we simulate the shadow cast by the same. Besides, we also plot the redshift function using the best-fit parameter values and fit it with the data.
 
    \item In conclusion, the observed flat rotation curves and the shadow radius can be used to determine the spacetime geometry of the dark matter-dominated galaxy. Apart from explaining the dynamics of the galaxy away from the galactic center, it can also provide insight into the nature of the compact object present at the galactic center. It should also be noted that we do not claim that the proposed spacetime satisfies all the properties of a galaxy. The proposed spacetime is a simple model of a galactic spacetime that is constructed by using some of the important characteristics of a galaxy.
 \end{itemize}

%
\end{document}